\documentclass[a4paper,10pt]{article}

\input{epsf.sty}

\begin{document}

\title{
 TIME-SYMMETRIZED  QUANTUM THEORY}
\author{ Lev Vaidman}
\date{}
\maketitle

\begin{center}
{\small \em School of Physics and Astronomy \\
Raymond and Beverly Sackler Faculty of Exact Sciences \\
Tel Aviv University, Tel-Aviv 69978, Israel. \\}
\end{center}

\vspace{.1cm}
\begin{abstract}
A brief review of the time-symmetrized quantum formalism originated by
Aharonov, Bergmann and Lebowitz is presented. Symmetry of various
measurements under the time reversal is analyzed. Time-symmetrized
counterfactuals are introduced. It is argued that the time-symmetrized
formalism  demonstrates  novel profound features of quantum theory and that
recent criticism of the formalism is unfounded.
\end{abstract}


\vskip .2cm
 \noindent
{\bf 1. Introduction.~~ }
\vskip .1cm

The time-symmetrized quantum theory (TSQT) originated in a seminal
work of Aharonov, Bergmann, and Lebowitz (ABL) \cite{ABL}. Since then
Aharonov and  co-workers have developed a rich formalism \cite{AV91}
which has led to the discovery of  numerous bizarre effects in quantum theory
\cite{AACV,s-xyz,s-100,t-m,K<0}. Alternative time-symmetrized
approaches have  been suggested by other authors \cite{Cr,Gr,GeHa}. Recently, however,
the validity of some of the results of the TSQT, especially in the
context of {\it counterfactual} interpretation of the ABL rule, have
been questioned \cite{SS}. The purpose of this paper is to explain the
meaning of the TSQT developed by Aharonov's group, to review
time-symmetry properties in the framework of this formalism, to give a
brief answer to the critics of the TSQT and to define time-symmetrized
counterfactuals in quantum theory.

The novelty of the TSQT follows from the observation that in quantum theory, contrary to
the classical physics, the future measurements might add
information about the present of a  system.  In the standard
approach, a quantum system at a given time $t$ is described completely
by a quantum state defined by the results of measurements performed on
the system in the past (relative to the time $t$).  In the TSQT a
quantum system at a given time $t$ is described by a {\it two-state
  vector} defined by the results of measurements performed on the
system in the past and in the future (relative to the time $t$). Thus,
while standard quantum theory deals with pre-selected  systems,
the TSQT analyzes pre- and post-selected  systems.

The purpose of the description of a quantum system by quantum state or
by two-state vector is to connect the class of {\it preparations}
which lead to the same set of effects on other systems. In Section 2 I
shall explain how to prepare a quantum system described by various
descriptions. In section 3 I analyze the outcomes of {\em ideal
  measurements} performed on systems prepared in different ways.
Section 4 is devoted to {\em generalized ideal measurements} recently
introduced by Shimony \cite{SHI}. Section 5 analyzes the {\em
  counterfactual interpretation} of the ABL rule. In section 6 I
describe {\em weak measurements} \cite{AV90}. Section 7 concludes the
paper with a brief discussion of the connection between the TSQT and the
many-worlds interpretation \cite{MWI} of quantum theory.

\vskip .6 cm \noindent
{\bf  2. Preparation of a Quantum System.~~}
\vskip .2 cm

\noindent
{\bf (a)} {\it Pre-selected quantum systems.}

In standard quantum theory a complete description of a system at a
given time $t$ is given by a forward evolving  quantum
state, see  Fig. 1{\it a}:
\begin{equation}
 \label{ket}
 |\Psi \rangle.
\end{equation}
In order to prepare the quantum state (\ref{ket}) we have to perform a
complete measurement in the past time $t_1$, $t_1<t$, and obtain a
specific outcome $A=a$ such that after the unitary evolution from
$t_1$ to $t$ the system will be in the desired state:
\begin{equation}
  \label{ket=}
 |\Psi \rangle ~=~ U(t_1, t) ~|A=a\rangle .
\end{equation}

\noindent
{\bf (b)} {\it Pre- and post-selected quantum systems.}
\vskip .1 cm
The basic description of a quantum system in the TSQT is given by a two-state
vector, see Fig. 1{\it b}:
\begin{equation}
  \label{2sv}
 \langle\Phi |~ |\Psi \rangle  .
\end{equation}
In order to prepare the two-state
vector (\ref{2sv}),  
in addition to the   measurement $A=a$ at
  time $t_1$, we have also to perform a complete measurement at $t_2$,
 $t_2>t$, and obtain
\vskip .5cm
\epsfxsize=12cm
 \centerline{\epsfbox{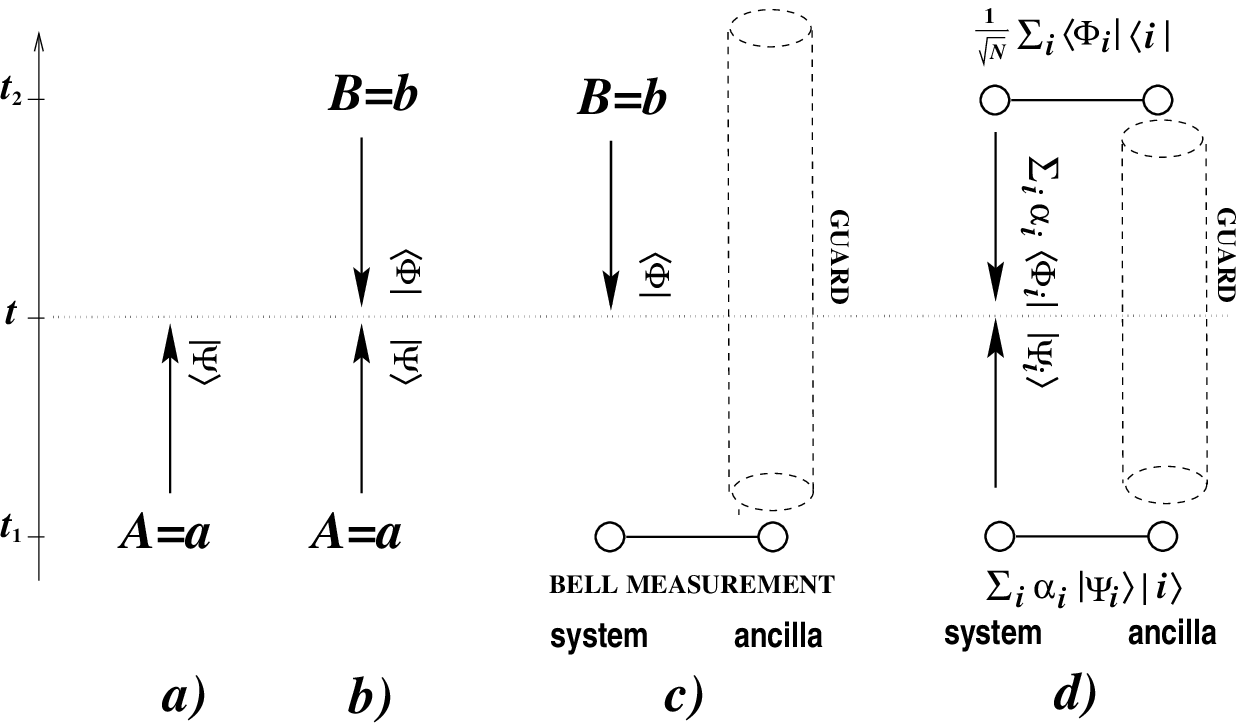}}
\vskip .2cm
\noindent
{\bf Fig. 1 Description of quantum systems:}~
 (a) pre-selected, (b) pre- and
post-selected, (c) post-selected, and (d) generalized pre- and
post-selected.

\noindent
  a
specific outcome $B=b$  such that  the backward time  evolution  from $t_2$
to $t$ will yield the desire state
\begin{equation}
  \label{bra=}
\langle\Phi | ~=~\langle B=b|~ U(t_2, t) .
\end{equation}

In this presentation there is complete symmetry between preparation of
the  states  $\langle\Phi |$ and $ |\Psi \rangle $ which constitute
the two-state vector: measurement of $A$ at $t_1$ leads to 
$ |\Psi \rangle $ and the measurement of $B$ leads to   $\langle\Phi
|$. Both measurements might not yield the desired outcomes, so we need
several systems out of which we  pre- and post-select the one
which is described by the two-state vector (\ref{2sv}). However, this
is only an apparent
symmetry. There is an intrinsic difference in preparation of $ |\Psi
\rangle $ and   $\langle\Phi |$. For preparation of  $ |\Psi \rangle $
a single system is enough. If the measurement of $A$ yields a
different outcome $a'$ we can perform a fast unitary operation which will
 change 
$|A=a'\rangle$ to $|A=a\rangle$ and then the time evolution to time
$t$ will bring the system to the state  $ |\Psi \rangle $. This
procedure is impossible for preparation of the backward evolving
state   $\langle\Phi |$. Indeed, if the outcome of the measurement of  $B$
does not yield  $b$, we cannot read it and then make an appropriate unitary
operation {\it before} $t_2$ in order to get the state  $\langle\Phi
|$ at time $t$. We need several systems to post-select the desired
result (unless by chance the first system  has the desired outcome).

In this paper I shall analyze symmetry under the interchange
$\langle\Phi|~|\Psi \rangle ~\leftrightarrow ~ \langle \Psi|~|\Phi
\rangle$.  This will be considered as a symmetry under reversal of the
direction of the arrow of time.  It is important to note that in
general this interchange is not equivalent to the interchange of  the
measurements $A=a$ and  $B=b$.
 (An example showing the non-equivalence can be found in the Appendix
of Shimony's paper \cite{SHI}.)  However, in order to simplify the
discussion, I will assume that the free Hamiltonian is zero, and
therefore $|\Psi \rangle ~=~ |A=a\rangle$ and $\langle\Phi |
~=~\langle B=b|$. In this case, of course, the reversal of time arrow
is identical to the interchange of the measurements at $t_1$ and
$t_2$.

\vskip .1 cm
\noindent
{\bf (c)} {\it Post-selected quantum systems.}
\vskip .1 cm

I have presented above a description of quantum system by  a single forward-evolving
quantum state (\ref{ket}) and by a two-state vector  (\ref{2sv}). It is
natural to ask: Are there  systems described by a  single backward-evolving
quantum state? The notation for such a state is 
\begin{equation}
\label{bra}
 \langle\Phi |.
\end{equation}
A measurement of $B$ at time $t_2$, even in the case it yields the
desired outcome $B=b$, is not enough. The difference between
preparation of (\ref{ket}) and (\ref{bra}) is that at present, $t$, the
future of a quantum system does not exist (the future measurements
have not been performed yet), but the past of a quantum system exists:
even if {\it we} do not know it, there is a  quantum state
of the system evolving towards the future  defined by the
results of  measurements in the past. Therefore, in order to prepare a
quantum system described by a backward evolving quantum state
(\ref{bra}), in addition to the post-selection measurement performed
after  time $t$, we have to {\it erase} the past.
 (We do not have to perform a special erasure procedure for
preparation of the pre- and post-selected system described by a
two-state vector (\ref{2sv}) because the complete measurement of $A$
at $t_1$ erases all prior information.)

In order to erase the past of a quantum system we can use another
quantum system, {\it an ancilla}.  The idea is to correlate an {\it
  unknown} future  of the ancilla with the past of our system. For
instance, such a correlation 
can be achieved using  a Bell-type
measurement, see Fig. 1{\it c}. For a spin-1/2 particle it can be realized by a sequence of two two-particle
measurements \cite{AAV86}: 
\begin{equation}
    \label{bell}
  (\sigma_x + \sigma_x^{(an)}){\rm mod} 4, ~~~~
 \label{bell1} (\sigma_z + \sigma_z^{(an)}){\rm mod} 4 .
\end{equation}
Each measurement has two possible outcomes, 0 or 2. The four possible
combinations   correspond to four possible Bell state. Each Bell
state yields complete correlation between the states of the two
systems. A generalization to a system with continuous degrees of freedom can be
done similarly to the generalization of {\it teleportation} to such
systems \cite{tel-V}. The erasure procedure has to be supplemented by
``guarding  the ancilla'' against any possible measurements after the Bell-type measurement. This will ensure that the
past of the system will be correlated to the unknown future of the ancilla.

\vskip .1 cm
\noindent
{\bf (d)} {\it Generalized pre- and post-selected quantum systems.}
\vskip .1 cm

The technique of ``guarded ancilla'' is also used to create a quantum
system described by  a {\it generalized two-state vector}
\begin{equation}
  \label{g2sv}
 \sum_i  \alpha_i  \langle\Phi_i |~ |\Psi_i \rangle  
\end{equation}
In order to prepare the generalized two-state vector (\ref{g2sv}) we
have to prepare at $t_1$ the system and the ancilla in a correlated
state $ \sum_i \alpha_i |\Psi_i \rangle |i \rangle$, where $\{
|i \rangle\}$ is a set of orthonormal states of the ancilla.
Then we have to ``guard'' the ancilla such that there will be no  measurements or any
other interactions performed on the ancilla  until the post-selection
measurement of a projection on the correlated state $1/\sqrt N \sum_i~
 |\Phi_i \rangle |i \rangle $, see Fig. 1{\it d}. If we obtain the desired outcome then the system is
described at time $t$ by the generalized two-state vector (\ref{g2sv}).

In the standard quantum theory the most general description of a
quantum system is given by a {\it mixed} state or {\it density
  matrix}. For example, a quantum system after the ``erasing the
past'' procedure proposed above is described by a density matrix and
cannot be described by a pure state.  It is believed, however, that
there is always a larger system (which includes the quantum system) in
a pure quantum state. (In particular, the composite system including
the system with ``erased past'' and the ancilla which was used in the
erasure procedure is in a pure state.) There is certain similarity
between this situation and a system described by a generalized
two-state vector: the composite system which includes the ancilla is
described by a two-state vector (\ref{2sv}). However, the analogy is
not exact. A generalized two-state vector is not the most general
description of a quantum system in the TSQT formalism -- it is the most
general {\it complete} description. It is possible to define the
generalization of a two-state vector to a {\it mixed} case when
various two-state vectors are correlated to another ancilla (wich is
not post-selected). Although the generalization is
straightforward, it is not obvious  what is its most convenient
form. For a powerful, but somewhat cumbersome formalism see Ref.
\cite{RA}.

\vskip .6 cm \noindent
{\bf  3. Ideal  Quantum Measurements.~~}
\vskip .2 cm

In this section I shall discuss how a quantum system characterized by
a certain description interacts with other systems. Some particular
types of interactions are named  {\it measurements}  and the effect of
these interactions characterized as the results of these
measurements. The basic concept is an  {\it ideal quantum measurement}
of a an observable $C$. This operation is defined for pre-selected
quantum systems in the following way: 
\begin{quotation}
 {\bf (i)} If the  state of a quantum system before the measurement was
an eigenstate of $C$ with an eigenvalue $c_n$ then the outcome of the
measurement is $c_n$ and the quantum state of the system is not changed.
\end{quotation}
The standard implementation of the ideal quantum measurement  is modeled by the von Neumann
 Hamiltonian \cite{neumann}: 
\begin{equation}
  \label{neumann}
 H = g(t) p C,
\end{equation}
where $p$ is the momentum conjugate to the pointer variable $q$, and
the normalized coupling function $g(t)$ specifies the time of the
measurement interaction. The outcome of the measurement is the shift
of the pointer variable during the  interaction. In the
ideal measurement the function $g(t)$ is nonzero only during a very short
period of time, and the free Hamiltonian  during this period of time can be neglected.

 For a quantum system described by the two-state
vector (\ref{2sv}) the probability for an  outcome $c_n$ of an ideal
measurement of an  observable $C$ is given by
\cite{ABL,AV91}
\begin{equation}
  \label{ABL}
 {\rm Prob}(c_n) = {{|\langle \Phi | {\bf P}_{C=c_n} | \Psi \rangle |^2}
\over{\sum_j|\langle \Phi | {\bf P}_{C=c_j} | \Psi \rangle |^2}} .
\end{equation}
This formula is explicitly time-symmetric: First, both $\langle\Phi |$
and $|\Psi \rangle$ enter the equation on equal footing. Second, the
probability (\ref{ABL}) is unchanged under the interchange
$\langle\Phi|~|\Psi \rangle~ \leftrightarrow ~\langle \Psi|~|\Phi \rangle$.

  For a quantum system described by a {\it generalized two-state
vector} (\ref{g2sv}) the probability for an  outcome $c_n$ is given by
\cite{AV91}
\begin{equation}
  \label{ABL-gen}
 {\rm Prob}(c_n) = {{|\sum_i \alpha_i \langle \Phi_i | {\bf P}_{C=c_n} | \Psi_i \rangle |^2}
\over{\sum_j|\sum_i \alpha_i \langle \Phi_i | {\bf P}_{C=c_j} | \Psi_i \rangle |^2}} .
\end{equation}
This formula is also  time-symmetric. Indeed,   
 $\langle\Phi_i |$ and ~$|\Psi_i \rangle$ enter the equation on equal
 footing. The manifestation of the symmetry of this formula under the
 reversal of the arrow of time includes complex conjugation of the
 coefficients. The
 probability (\ref{ABL-gen})  is unchanged  under the interchange
$\sum_i \alpha_i \langle \Phi_i | ~ | \Psi_i \rangle ~ \leftrightarrow
~ \sum_i \alpha_i^{\ast} \langle \Psi_i | ~ | \Phi_i \rangle$.

Another important generalization of the formula (\ref{ABL}) is for the
case in which the post-selection measurement is not complete and
therefore it does not specify a single post-selection state $\langle
\Phi|$. Such an example was recently considered by Cohen \cite{CO} in
(an unsuccessful \cite{CO-co}) attempt to find constraints to the applicability of
the ABL formula. In this case, the post-section measurement is a projection on a
{\em degenerate} eigenvalue of an observable $B=b$. The modified ABL formula
is \cite{CO-co}:
\begin{equation}
  \label{ABL-new}
 {\rm Prob}(c_n) = {{\Vert {\bf P}_{B=b}  {\bf P}_{C=c_n} | \Psi \rangle \Vert^2}
\over{\sum_j \Vert {\bf P}_{B=b}  {\bf P}_{C=c_j} | \Psi \rangle \Vert^2}} .
\end{equation}
This form of the ABL formula allows to connect it to the standard
formalism of quantum theory in which there is no post-selection. In the
limiting case when the projection operator  ${\bf P}_{B=b}$ is just
unity operator {\bf I}, we  obtain the usual expression:
\begin{equation}
  \label{ABL-pre}
 {\rm Prob}(c_n) = ||  {\bf P}_{C=c_n} | \Psi \rangle ||^2 .
\end{equation}

\vskip .6 cm \noindent
{\bf  4. Generalized Ideal  Quantum Measurements.~~}
\vskip .2 cm
 Recently,   Shimony \cite{SHI}  proposed  the following   generalization of
the concept of ideal measurements by weakening the requirement given
in definition (i):
\begin{quotation}
 {\bf (ii)} If the  state of a quantum system before the measurement is
an eigenstate of $C$  with an eigenvalue $c_n$ then the outcome of the
measurement is $c_n$ and the quantum state of the system after the
measurement  remains in the subspace  of the eigenstates
corresponding to this  eigenvalue.
\end{quotation}
The difference between  definitions (i) and (ii) is that while in
the ideal
measurements the change of the state during the measurement happens
only if the initial state was a superposition of states with different
eigenvalues, the generalized ideal measurement procedure changes
also the eigenstates themselves. (This might happen only for
degenerate eigenvalues.) Shimony has constructed a generalized ideal
measurement for which the ABL formula (\ref{ABL}) does not hold. I
shall derive Shimony's result using another example.

Consider a composite system of two spin-$1\over 2$ particles. The
variable to be measured is the value of the total spin of the system
which can be 0 or 1. The total spin 0 corresponds to a nondegenerate
state (singlet), while total spin 1 corresponds to three orthogonal
eigenstates (triplet). The generalized ideal measurement do not
change measured value (the total spin), but it might change a commuting
variable. In this example the measurement  changes cyclicly the total $\hat z$ component
of the spin (for triplet). This measurement is described by the
following unitary transformation:
\begin{eqnarray}
  \label{shi1}
\nonumber\mid 0,0\rangle    |R\rangle_{MD} &\rightarrow& \mid 0,0\rangle  |0\rangle_{MD}\\
\mid 1,-1\rangle |R\rangle_{MD} &\rightarrow& \mid 1,0\rangle |1\rangle_{MD}\\
\nonumber\mid 1,0\rangle |R\rangle_{MD} &\rightarrow& \mid 1,1\rangle |1\rangle_{MD}\\
\nonumber\mid 1,1\rangle |R\rangle_{MD} &\rightarrow& \mid 1,-1\rangle  |1\rangle_{MD}
\end{eqnarray}
where $\mid 0,0\rangle$,  $\mid 1,-1\rangle$ etc. denote the states
of the composite system in $S^2$, $S_z$ representation, while $
|R\rangle_{MD}$, $ |0\rangle_{MD}$ and $ |1\rangle_{MD}$ denote the
``ready'' and ``0'' and ``1'' final states of the measuring device.  

Now consider the measurement of the total spin performed on a system
pre- and post-selected in a two-state vector
\begin{equation}
  \label{psi1-2}
\langle \Phi |~  |\Psi\rangle = {1\over 2 } ~(\langle 0,0|  + 
\langle 1,1|)~ (|0,0\rangle + |1,0\rangle).
\end{equation}
Given the pre-selection, $  |\Psi\rangle = 1/{\sqrt 2}  ~(|0,0\rangle + |1,0\rangle)$,
 only, the probability for the two possible
outcomes are equal:
\begin{equation}
  \label{p(|psi)}
    {\rm Prob}(0 ;  ~ |\Psi\rangle)=  {\rm Prob}(1 ;~ |\Psi\rangle)=
  {1\over 2}.
\end{equation}
If the outcome is ``0'' then, after the measurement, the state of the
system is $|0,0\rangle$;
if,  the outcome is ``1'' then the final state is  $|1,1\rangle$.
It turns out that in both cases  the probability for the
post-selection of the state, $\langle \Phi |= 1/\sqrt 2  ~(\langle 0,0|  + 
\langle 1,1|)$, is:
\begin{equation}
  \label{p(phi)}
  {\rm Prob}(\langle \Phi|~; 0)  =  {\rm Prob}(\langle \Phi| ~; 1)  =  {1\over 2}.
\end{equation}
Then,  Bayes theorem yields  the probability of
obtaining the outcome ``1''
in the generalized measurement performed on the
pre- and post-selected system:
\begin{equation}
  \label{bay}
  {\rm Prob}(1 ;~ \langle \Phi| ~|\Psi\rangle)        =  {{
 {\rm Prob}(1 ;~ |\Psi\rangle) {\rm Prob}(\langle \Phi | ~; 1)}\over
 {{\rm Prob}(0 ;~ |\Psi\rangle) {\rm Prob}(\langle \Phi |~; 0) +
 {\rm Prob}(1 ;~ |\Psi\rangle) {\rm Prob}(\langle \Phi |~; 1)}}
 ={1\over 2}
\end{equation}
Let us repeat the calculation for the time reversed two-state vector:
\begin{equation}
  \label{psi1-2'}
  \langle \Psi  |~  |\Phi \rangle = {1\over 2} (\langle 0,0|  + \langle 1,0|)~ (|0,0\rangle + |1,1\rangle).
\end{equation}
Now, after the generalized measurement with outcome ``1'', the state of
the system must be $|1,-1\rangle$. This state, however, is orthogonal
to the post-selected state  and therefore  this outcome is
impossible.
Thus, ${\rm Prob}(\langle \Psi |
~; 1)= 0$ and consequently, 
 ${\rm Prob}(1 ;~ \langle \Psi| ~|\Phi\rangle) =0$. 

I have shown that in this example there is an asymmetry:~ ${\rm Prob}(1 ;~ \langle \Phi| ~|\Psi\rangle) \neq
 {\rm Prob}(1 ;~ \langle \Psi| ~|\Phi\rangle)$. Therefore,  the ABL
 formula, which is symmetric under such interchange, cannot hold. (It yields $ {\rm Prob}(1 ;~ \langle \Phi| ~|\Psi\rangle) =
 {\rm Prob}(1 ;~ \langle \Psi| ~|\Phi\rangle)= 0. $)
 I find this result not too disturbing since  these measurements
have an unusual property even in a situation in which the ABL formula
is not involved: generalized ideal measurements  of commuting observables
might disturb each other.

\vskip .6 cm \noindent
{\bf  5. Counterfactual Interpretation of the ABL Rule.~~}
\vskip .2 cm

Several authors  criticized the TSQT  
because of the alleged  conflict between counterfactual interpretations of the ABL rule
and predictions of quantum theory \cite{SS,CO,MI}.
The  form of all these inconsistency proofs is as follows:
The probability of an outcome $C=c_n$ of a quantum measurement performed on a
pre-selected system, given correctly by (\ref{ABL-pre}), is
considered. In order to allow the analysis 
using the ABL formula, a measurement at a later time, $t_2$, with two
possible outcomes, which we denote by  ``$1_f$'' and ``$2_f$'',
 is introduced. 
The suggested application of the ABL rule is expressed in the formula
for the
probability of the result $C=c_n$:
\begin{equation}
  \label{p1}
  {\rm Prob}(C=c_n)~=~  {\rm Prob}(1_f) ~ {\rm Prob}(C=c_n~; 1_f) +
  {\rm Prob}(2_f) ~{\rm Prob}(C=c_n ~; 2_f),
\end{equation}
where $ {\rm Prob}(C=c_n ~; 1_f)$ and $ {\rm Prob}(C=c_n~ ; 2_f)$ are the 
conditional probabilities given by the ABL formula, (\ref{ABL}), and $ {\rm
  Prob}(1_f)$ and $ {\rm Prob}(2_f)$ are the probabilities of the
results of the final measurement.
In the proofs, the authors show that Eq. (\ref{p1}) is not valid and conclude
that  the ABL formula is not applicable to this example and 
therefore  it is not applicable in general.

I have argued \cite{DTSQT,TSCF,CO-co} that the error in calculating
equality (\ref{p1}) does not arise from the conditional probabilities
given by the ABL formula, but from the calculation of the
probabilities $ {\rm Prob}(1_f)$ and $ {\rm Prob}(2_f)$ of the final
measurement. In all the three alleged proofs the probabilities $ {\rm
  Prob}(1_f)$ and $ {\rm Prob}(2_f)$ were calculated on the assumption
that {\rm no} measurement took place at time $t$. Clearly, one cannot
make this assumption here since then the discussion about the
probability of the result of the measurement at time $t$ is
meaningless. Thus, it is not surprising that the value of the
probability ${\rm Prob}(C=c_n)$ obtained in this way comes out
different from the value predicted by the quantum theory.
Straightforward calculations show that the formula (\ref{p1}) with the
probabilities $ {\rm Prob}(1_f)$ and $ {\rm Prob}(2_f)$ calculated on
the condition that the intermediate measurement has been performed
leads to the result predicted by the standard formalism of quantum
theory.
 
The analysis of counterfactual statements considers both {\em actual}
and {\em counterfactual} worlds. The statement is considered to be
true if it is true in counterfactual worlds ``closest'' to the actual
world.  In the context of the ABL formula, in the actual world the
pre-selection and the post-selection has been successfully performed,
but the measurement of $C$ has not (necessarily) been performed. On
the other hand, in counterfactual worlds the measurement of $C$ has
been performed.  The problem is to find counterfactual worlds
``closest'' to the actual world in which the measurement of $C$ has
been performed. The fallacy in all the inconsistency proofs is that
their authors have considered counterfactual worlds in which $C$ has
not been measured.
 
Even if we disregard  this fallacy  there is still a difficulty in defining the
``closest'' worlds in the framework of the TSQT. In standard quantum
theory it is possible to use the most natural definition of the
``closest'' world. Since the future is considered to be irrelevant for
measurements at present, $t$, only the period of time before $t$
is considered. Then the definition is: 
\begin{quotation}
{\bf (iii)} Closest counterfactual  worlds are  the
worlds in which the system is described by the same quantum state as
in the actual world.
 \end{quotation}
In the framework of the TSQT, however, this definition is
not acceptable. In the time-symmetric approach the period of time before and
after $t$  is considered. The measurement of $C$
constrains the possible states immediately after $t$ to the
eigenstates of $C$. Therefore,  if in the actual world the state immediately  after
$t$ is  not an
eigenstate of $C$, no counterfactual world with the same state
exists. Moreover, there is the same problem with the backward evolving
quantum state (the concept which does not exist in the standard
approach) in the period of time before  $t$. I proposed \cite{TSCF} to
solve this  difficulty by adopting the following definition of the
closest world:
\begin{quotation}
{\bf (iv)}  Closest counterfactual  worlds are  the
worlds in which the results of all measurements performed on the system
(except the measurement at time $t$) are the same as in 
 actual world.
 \end{quotation}
 For the
 pre-selected only situation, this definition is equivalent to  (iii), but it is also applicable to the
symmetric pre- and post-selected situation.

An important example of  counterfactuals in quantum
theory are  ``elements of reality''.
I have proposed
  a modification of the definition of elements of reality  applicable
  to the framework of the TSQT \cite{V93}:
 \begin{quotation}
 {\bf (vi)~}   If we can {\em infer} with certainty that the result of
   measuring at time $t$ of an observable $C$ is $c$, then, at time
   $t$, there exists an element of reality $C=c$.
\end{quotation}
The word ``infer'' is neutral relative to past and future. The
inference about results at time $t$ is based on the results of
measurements on the system performed both before and after time $t$.
Note, that there are  situations (see example with a particle in three
boxes below) in which we can ``infer'' some facts that cannot be
obtained by neither ``prediction'' based on the past results nor 
``retrodiction'' based on the future results separately.

\vskip .6 cm \noindent
{\bf  6. Weak  Measurements.~~}
\vskip .2 cm

The most interesting phenomena which can be seen in the framework of
the TSQT  are related to {\it weak measurements} \cite{AV90}. A weak
measurement is a standard measuring procedure (described by the
Hamiltonian (\ref{neumann}) with weakened coupling. In ideal
measurement the initial position of the pointer $q$ is well localized
around zero and therefore the conjugate momentum $p$ has a very large
uncertainty which leads to a very large uncertain Hamiltonian  of the
measurement (\ref{neumann}). In weak measurement, the initial state of
the measuring device is such that $p$ is localized around zero with
small uncertainty. This leads, of course, to a large uncertainty in $q$ and
therefore  the measurement becomes imprecise.  However,
by performing the weak measurement on an ensemble of $N$ identical systems
we improve the precision by the factor of $\sqrt N$ and in some
special cases
we can obtain  good precision even in a measurement performed on a single
system \cite{AACV}. The outcome of a weak measurement of a variable $C$
is the {\it weak value}. The weak value of a variable $C$ of a system
described by the two-state vector  $\langle\Phi |~ |\Psi \rangle$ is: 
\begin{equation}
C_w \equiv { \langle{\Phi} \vert C \vert\Psi\rangle 
\over \langle{\Phi}\vert{\Psi}\rangle } .
\label{wv}
\end{equation}
Strictly  speaking, the readings of the pointer of the measuring device
will cluster around Re$(C_w)$. In order to find Im$(C_w)$  one should measure the shift in $p$ \cite{AV90}.

The weak value is symmetric under the interchange  $\langle\Phi |~
|\Psi \rangle ~ \leftrightarrow ~ \langle\Psi |~ |\Phi \rangle $ provided we
perform complex conjugation of the weak value together with the
interchange. This is similar to complex conjugation of the Schr\"odinger
wave function under the time reversal. Thus, also for weak measurement
there is time reversal symmetry: both  $\langle\Phi |$ and  $|\Psi
\rangle$ enter the formula of the weak value on the same footing and
there is symmetry under the interchange of the pre- and post-selected states.

The weak value for a system described by a generalized two-state vector
(\ref{g2sv}) is \cite{AV91}:
\begin{equation}
  \label{wv-gen}
 C_w = {{\sum_i \alpha_i \langle \Phi_i | C | \Psi_i \rangle }
\over{\sum_i \alpha_i \langle \Phi_i  | \Psi_i \rangle }} .
\end{equation}
This expression  is also  symmetric under the time reversal,
i.e., the interchange \break
$\sum_i \alpha_i \langle \Phi_i | ~ | \Psi_i \rangle  ~ \leftrightarrow
~ \sum_i \alpha_i^{\ast} \langle \Psi_i | ~ | \Phi_i \rangle~$ leads to
$~C_w  \leftrightarrow C_w^\ast$.

Next, let us look at the  expression for the weak value when the
post-selection measurement is not complete. Consider a system
pre-selected 
in the state  $|\Psi\rangle$ and post-selected by the the
measurement of degenerate eigenvalue $b$ of a variable  $B$.  (The ABL
formula for the probability of $C=c_n$ for such situation  is given by (\ref{ABL-new}).
\begin{equation}
  \label{wv-new}
 C_w = {{ \langle{\Psi} \vert {\bf P}_{B=b}  C \vert \Psi \rangle} 
\over {\langle \Psi {\bf P}_{B=b}\vert \Psi \rangle }} .
\end{equation}

This formula allows us to find the outcome of a weak measurement
performed on a pre-selected (only) system. Replacing ${\bf P}_{B=b}$ by
the unity operator yields the result that the weak value of a
pre-selected system in the state $|\Psi\rangle$ is the expectation
value:
\begin{equation}
  \label{wv-exp}
 C_w =  \langle{\Psi} \vert   C \vert\Psi\rangle . 
\end{equation}

 Weak values have
many interesting properties, in particular, $(A+B)_w= A_w+ B_w$, even
for non-commuting observables $A$ and $B$.  I have considered weak
values as {\em weak-measurement of reality} \cite{wmer}, but the weak value
is not just a theoretical concept related to a gedanken experiment.
Recently, weak values have been measured in a real
laboratory \cite{w-exp}.

An  interesting connection between weak and strong
(ideal) measurements is a theorem \cite{AV91} which
says that if the probability for a certain value to be the result of a
strong measurement is 1, then the corresponding weak measurement must also
yield the same value, i.e., element of reality $C=c$, implies
weak-measurement element of reality $C_w=c$. 
Consider a single particle located in three separated boxes $A, B$, and
$C$ pre- and post- selected in the two-state
vector
\begin{equation}
\langle\Phi|~|\Psi\rangle = 
{{1\over 
  3}}(\langle A|+ \langle B|-\langle C|)~(|A\rangle + |B\rangle +
|C\rangle).
\end{equation}
A set of counterfactual statements for this particle is: 
\begin{equation} 
\label{3b-str}
{\rm \bf P}_A = 1 ,~~~~
{\rm \bf P}_B = 1 ,~~~~
{\rm \bf P}_A + {\rm \bf P}_B + {\rm \bf P}_C = 1 .
\end{equation}
Or, in words: if we open box $A$, we find the particle there for sure;
if we open box $B$ (instead), we also find the particle there for
sure; if we open simultaneously all boxes, we find the particle in one
of them for sure. These counterfactual statements lead us to
statements about weak-measurement elements of reality:
\begin{equation} 
({\rm \bf P}_A)_w = 1 ,~~~~
({\rm \bf P}_B)_w = 1 ,~~~~
({\rm \bf P}_A + {\rm \bf P}_B + {\rm \bf P}_C)_w = 1 .
\end{equation}
From these results we can also deduce that $({\rm \bf P}_C)_w = -1$.
Therefore, for every sufficiently weak interaction the effective
coupling to this single particle is equivalent to the weak coupling to
a single particle in box $A$, a single particle in box $B$ and {\em
  minus} one particle in box $C$. The meaning of the latter is that
for a pre- and post-selected ensemble of many such particles there is
an effective {\em negative} pressure in box $C$. However, an experiment which
will test this is very difficult because the probability to obtain
such an ensemble is very low.

\vskip .6 cm \noindent
{\bf 7. Conclusions.~~}
\vskip .2cm

In this paper I presented basic concepts of the TSQT. In order to get
a complete picture one should read also other works which were cited
here, but some topics were analyzed in this paper more completely than
in other publications. Among them the issue of preparation of a
quantum system  described in a time-symmetrized way. Here ``pre'' of
the 
word {\it pre}paration is somewhat misleading, but I could not find
more appropriate word. The absence of a word in English which
describes exactly the situation follows
from the fact that until very recently we had no reason to believe
that the result of future measurements are relevant to the discussion
of a measurement at present.  In this context I want to mention that
the time-symmetrized quantum theory fits well into the many-worlds
interpretation (MWI) \cite{MWI}, my preferred interpretation of
quantum theory \cite{my-MWI}.  The counterfactual worlds corresponding
to different outcomes of quantum measurements have in the MWI an
especially clear meaning: these are subjectively actual different
worlds.  In each world the observers of the quantum measurement call
their world the actual one, but, if they believe in the MWI they have no
paradoxes about ontology of the other worlds. The apparent paradox
that a weak value at a given time might change from an expectation
value to a weak value corresponding to a particular post-selection is
solved in a natural way: in a world with pre-selection only (before the
post-selection) the weak value is the expectation value; then this
world splits into several worlds according to results of the
post-selection measurement and in each of these worlds the weak value
will be that corresponding to the particular post-selection.

Another issue which was  elaborated here in detail is the time-symmetry of the
generalized ideal measurements recently proposed by Shimony. I showed
explicitly why these measurements are not symmetric under the time reversal.

An important novel issue which was briefly considered here  is
time-symmetrized counterfactuals in quantum theory. These results have
not been published yet in other journals, but one can read more on this subject in the
preprints \cite{CO-co,DTSQT,TSCF}.

The most important issue in the framework of the TSQT, the issue of
weak measurements, was also discussed here only briefly and I recommend reading
the review lecture \cite{weak}. The connection to protective
measurements \cite{prot2} was not covered at all. The only novel point
related to weak measurements in this work is Eq. (\ref{wv-new}) which
yields the weak value in the case of partial post-selection.

It is a pleasure to thank Yakir Aharonov,
Jacob Grunhaus and
Stephen Wiesner for helpful discussions. The research was supported in
part by grant 614/95 of the Basic Research Foundation (administered by
the Israel Academy of Sciences and Humanities).


\begin{thebibliography}{99} 


\bibitem{ABL}
 Y. Aharonov, P. G. Bergmann  and J. L.
   Lebowitz,     Phys.  Rev.  {\bf B134},  1410 (1964).


\bibitem{AV91}
Y. Aharonov  and L. Vaidman, 
 J.  Phys. {\bf  A 24}, 2315 (1991).
\bibitem{AACV}
Y. Aharonov, D. Albert, A. Casher, and L. Vaidman,
{\it Phys. Lett.} {\bf  A 124}, 199 (1987).

\bibitem{s-xyz}
L. Vaidman, Y. Aharonov and D. Albert,
{\it Phys. Rev. Lett.} {\bf  58}, 1385 (1987).
 

\bibitem{s-100}
  Y. Aharonov,  D. Albert, and L. Vaidman,
{\it Phys. Rev. Lett.} {\bf  60}, 1351 (1988).


\bibitem{t-m}
 Y. Aharonov, J. Anandan, and L. Vaidman, {\it Phys. Rev.} {\bf
 A 47}, 4616 (1993).

\bibitem{K<0}
Y. Aharonov,  S. Popescu, D.~Rohrlich, and L.~Vaidman, {\it Phys.  Rev.}
{\bf A 48}, 4084 (1993).

\bibitem{Cr}
{\it Rev. Mod. Phys.} {\bf 58}, 647 (1986).

\bibitem{Gr}
R. B. Griffiths,
J. Stat. Phys. {\bf 36}, 219 (1984).
 
\bibitem{GeHa}
M. Gell-Mann   and J. B. Hartle,   
 in W. H. Zurek (ed), {\it Complexity, Entropy  and the Physics 
of Information}, (Reading: Edison-Wesley), pp. 425-459 (1990).


\bibitem{SS}
W. D. Sharp  and N. Shanks,
 Phil.  Sci. {\bf 60}, 488 (1993).

\bibitem{SHI}
A. Shimony,
{\em Erken.} {\bf 45}, 337 (1997).


\bibitem{AV90}
Y. Aharonov  and L. Vaidman, 
  Phys. Rev. {\bf  A 41}, 11 (1990).



\bibitem{MWI}
H. Everett,  {\it Rev. Mod.  Phys.}
  {\bf 29}, 454 (1957).


\bibitem{AAV86}
Y. Aharonov, D. Albert, and L. Vaidman,
{\it Phys. Rev.} {\bf  D 34}, 1805 (1986).
 

\bibitem{tel-V}
L. Vaidman,
{\it Phys. Rev.} {\bf A 49}, 1473 (1994).


\bibitem{RA}
B. Reznik and  Y. Aharonov,
Phys. Rev. A, 52, 2538 (1995).

\bibitem{neumann}
 J. von Neumann,  {\it Mathematical Foundations of Quantum Theory},
  (Princeton, University Press, New Jersey (1983).


\bibitem{CO}
O. Cohen. (1995), 
 Phys.  Rev. {\bf  A 51},  4373 (1995). 
 

\bibitem{CO-co}
L. Vaidman,
Tel-Aviv University preprint quant-ph/9703001.

\bibitem{MI}
D. J. Miller,
{\em Phys. Lett.} {\bf A 222}, 31 (1996).



\bibitem{DTSQT}
L. Vaidman,
Tel-Aviv University preprint quant-ph/9609007.
 

\bibitem{TSCF}
L. Vaidman,
Tel-Aviv University preprint TAUP 2459-97.

\bibitem{V93}
L. Vaidman,
{\em Phys. Rev. Lett.} {\bf 70}: 3369(1993).

\bibitem{wmer}
L. Vaidman,
 Found.  Phys. {\bf 26}, 895 (1996).


\bibitem{w-exp}  
 N.W.M. Ritchie, J.G. Story, and R.G. Hulet, {\it Phys Rev. Lett.} {\bf
66}, 1107 (1991).


\bibitem{my-MWI}
L. Vaidman,
 preprint quant-ph/9609006, to be published in {\it Int.
  Stud. Phil.  Sci.}
 
\bibitem{weak}
 L. Vaidman,
in  {\it Advances in Quantum Phenomena}, E. Beltrametti and J.M.
Levy-Leblond eds.,  NATO ASI Series B: Physics  Vol. 347, Plenum
Press, NY, pp. 357-373 (1995). 
 

\bibitem{prot2}
 Y. Aharonov and L. Vaidman, in {\it Potentiality, Entanglement and Passion-at-a-Distance},
R.S.Cohen, et al. (eds), Kluwer Academic Publishers, pp.1-8, (1997). 

\end{thebibliography}
\end{document}